\documentclass[aps,prl,twocolumn,10pt,groupedaddress,nofootinbib,superscriptaddress,notitlepage,showpacs,floatfix]{revtex4-1}
\usepackage{graphicx,graphics}
\usepackage{epstopdf}
\usepackage{subfigure}
\usepackage{dcolumn}   % needed for some tables
\usepackage{array}
\usepackage{amsthm}
\usepackage{amsmath}
\usepackage{amssymb}
\usepackage{amsfonts}
\usepackage{mathrsfs} %curled math symbols
\usepackage{bm}% bold math
\usepackage{braket} %Dirac notation
\usepackage{enumitem}
\usepackage{url}
\usepackage[pdfstartview=FitH,pdfencoding=auto]{hyperref}
\hypersetup{
    colorlinks=true,        		% false: boxed links; true: colored links
    linkcolor=blue,              	% color of internal links
    citecolor=blue,        			% color of links to bibliography
    filecolor=blue,          		% color of file links
    urlcolor=blue,              	% color of external links
    runcolor=cyan
			}
\usepackage[pdftex]{color}
\usepackage{multirow}
\usepackage{CJK}
\newcommand{\defeq}{:=}           % be defined as
\newcommand{\eqdef}{=:}           % be defined as

\newcommand{\ud}{\mathrm{d}}

\newcommand{\est}{\mathrm{est}}
\begin{document}
%==================================================================================================================================
% Title, authors, affiliation, dates, PACS.
%==================================================================================================================================
\begin{CJK*}{GB}{}
\title{Achieving Heisenberg limit under General Markov Noise without Ancilla}
\author{Yi Peng}
	\affiliation{Institute of Physics, Chinese Academy of Sciences, Beijing 100190, China}
	\affiliation{School of Physical Sciences, University of Chinese Academy of Sciences, Beijing 100190, China}
%------------------------------------------------------------------------------------------
\author{Heng Fan}
	\email{hfan@iphy.ac.cn}
	\affiliation{Institute of Physics, Chinese Academy of Sciences, Beijing 100190, China}
	\affiliation{CAS Center for Excellence in TQC, University of Chinese Academy of Sciences, Beijing 100049, China}
	\affiliation{Collaborative Innovation Center of Quantum Matter, Beijing 100190, China}
%------------------------------------------------------------------------------------------
\date{\today}
%=================================================================================================================================
% Abstract
%=================================================================================================================================
\begin{abstract}
	We consider probing qubits infested by general Makovian noise. It can be shown that one can restore Heisenberg limit (HL) via 
	full and fast control without any ancilla, as long as the Hamiltonian of the system has non-vanishing component perpendicular 
	to the noise.
\end{abstract}
\maketitle
\end{CJK*}
%=================================================================================================================================
% Main contents
%=================================================================================================================================
%==================================================================================================================================
% Introduction
%==================================================================================================================================
\emph{Introduction}---
	Quantum metrology is one of the three main streams in the thriving field of quantum information besides quantum communication 
	and quantum computation~\cite{usa2018NAS}. In the ideal noiseless condition, quantum resources such as entanglement can enhance 
	the estimation precision of parameters to breach the shot-noise limit (SNL) $1/\sqrt{NT}$ and reach the ultimate Heisenberg limit 
	(HL) $1/NT$, with $T$ being the total probing time and $N$ the number of probes 
	employed~\cite{giovannetti2004quantum-enhanced,giovannetti2006quantum,Demkowicz-Dobrzanski2015,Degen2017,Pezze2018,Kwon2019}. 
	The advantage provided by quantum resources ensures a promising future of applications as diverse as 
	spectroscopy~\cite{Bollinger1996}, accurate clock 
	construction~\cite{Kessler2014,Derevianko2011,Zhang2016}, gravitational wave detection~\cite{LIGO2013,Schnabel2010}, fundamental
	biology research and medicine development~\cite{Taylor2016}, and 
	{\it et. al}~\cite{giovannetti2004quantum-enhanced,Demkowicz-Dobrzanski2015,Degen2017,Pezze2018}. However, in the presence of 
	uncontrollable environmental noise, such an advantage would be diminished or destroyed in most 
	cases~\cite{Huelga1997,Escher2011general,Chin2012,Demkowicz-Dobrzanski2015,Demkowicz-Dobrzanski2012,Degen2017}. 
	As a result, extensive efforts have been made by many groups to reduce or even neutralize the detrimental effect of the noise,
	which include: a) searching for noise-resilient input states such as the spin-squeezed states and Dicke 
	states~\cite{sorensen2001many,Dunningham2002,ma2011quantum,Duan2011,Zhang2013,Pezze2013}, 
	b) designing schemes to harness the entanglement with noiseless ancilla systems~\cite{Demkowicz-Dobrzanski2014} or take
	advantage of the specific geometry of the noise~\cite{Chaves2013,Brask2015}, 
	c) designing noise-resilient control sequences via machine 
	learning~\cite{Hentschel2010,Hentschel2011,Lovett2013,Palittapongarnpim2017,Lumino2018},
	and d) adopting noise-suppression techniques from the fields of quantum computation and quantum communication such as 
	dynamical decoupling (DD)~\cite{Viola1999a,Viola1999b,Tan2013,Sekatski2016}, quantum error correction 
	(QEC)~\cite{Unden2016,Reiter2017,Sekatski2017a,Demkowicz-Dobrzanski2017,Zhou2018achieving,Layden2018,Layden2019,Cappellaro2019,
	nielsen2010quantum}
	and quantum teleportation~\cite{Matsuzaki2018}. Each routine has its own advantages. For DD and QEC which are among the most powerful
	tools developed for fault-tolerant quantum computation and communication, we have a good reason to expect their prowess in quantum
	metrology against environment noise. Their adoption to quantum metrology allows us to take advantage of the fruitful results 
	designated originally for practical quantum computation and communication~\cite{nielsen2010quantum}.

	Here we focus on the QEC scheme. In the presence of general Makovian noise, it has been shown by explicitly constructing QEC code 
	$C_\mathrm{O}$ such that HL can be achieved with the help of noiseless ancilla.
	This is possible if the probe Hamiltonian has non-vanishing component perpendicular to the Markovian noise. It is assumed that
	the ancilla systems are neither affected by the parameter field, for instance the magnetic field, nor any environment noise. If
	the ancilla systems suffer from Markovian noise also, one has to apply an \emph{inner code}
	$C_\mathrm{I}$ to obtain effectively noiseless ancilla systems. The HL can be achieved by 
	concatenating $C_\mathrm{I}$ and $C_\mathrm{O}$~\cite{Sekatski2017a,Demkowicz-Dobrzanski2017,Zhou2018achieving}. This kind of scheme is very suitable for the situation where there are at 
	least two types quantum systems of which one species has a much longer coherence time than others. A typical example is the 
	nitrogen-vacancy (NV) center where the decoherence rate of the nuclear spin is very small compared with that of the electron 
	spin~\cite{Maze2008,Liu2015,Unden2016}. In the case such as the superconducting qubit~\cite{Naghiloo2017} and trapped ion 
	system~\cite{Reiter2017} where there is only
	one kind of quantum system at disposal, the employment of the inner code $C_\mathrm{I}$ seems inevitable which can bring 
	further complexity. Devising an ancilla-free QEC scheme for quantum metrology is thus very important. Given 
	\emph{commuting Markovian noise}, it 
	has been further elaborated that there exists an ancilla-free QEC scheme ensuring the HL if part of the probe 
	Hamiltonian is perpendicular to the noise. The commuting Markovian noise covers common noises such as 
	the \emph{lossy bosonic \ channel} and \emph{dephasing}~\cite{Layden2019}. We generalize the result of 
	Ref.~\cite{Sekatski2017a,Demkowicz-Dobrzanski2017,Zhou2018achieving,Layden2018,Layden2019} by showing that the 
	HL can recovered by a QEC scheme for qubit probes affected by general Markovian noise without any ancilla. One can thus exorcise
	commuting as well as non-commuting noises for example the general \emph{Rank-one Pauli channel} which includes dephasing as
	well as bit flit channel~\cite{wilde2013quantum}.

	\emph{Quantum Metrology under general Markovian noise.}---Consider the quantum probe with Hamiltonian $\omega\hat{H}$ 
	suffering from Markovian noise described by the Lindblad operator $\hat{L}_k$ such that its state evolves
	homogeneously in time
	\begin{equation}
		\frac{\ud\hat{\rho}}{\ud{t}}
		=-i\omega\lbrack{\hat{H},\hat{\rho}}\rbrack
		+ \sum_{k=1}^R(\hat{L}_k\hat{\rho}\hat{L}_k^\dag - \frac{1}{2}\{\hat{L}_k^\dag\hat{L}_k,\hat{\rho}\}),
		\label{single_probe_selv}
	\end{equation}
	with $R$ being the rank of the noise. The information of both Hamiltonian $\hat{H}$ and Markovian noise is 
	available to us. Our task is to make an estimation $\omega_\est$ of $\omega$ as precisely
	as possible. It has been shown that such a time-homogeneous Markovian noise can render the estimation precision from HL to 
	SNL~\cite{Escher2011general,Demkowicz-Dobrzanski2012}. However, scuh a disastrous setback can be avoid. It has been shown 
	that~\cite{Sekatski2017a,Demkowicz-Dobrzanski2017,Zhou2018achieving,Layden2019}
	\begin{enumerate}
		\item given HNLS being true, HL can be restored by QEC with the help of noiseless ancilla systems;
		\item given commuting Markovian noise and HNLS being true, HL can be restored by QEC without any noiseless ancilla;
		\item if HNLS fails, HL can not be restored by QEC even with ancilla. 
	\end{enumerate}
	If the Hamiltonian 
	$\hat{H}$ can not be linearly spanned by the identity 
	operator $\openone$, Lindblad operators $\hat{L}_j$, $\hat{L}_j^\dag$ and their second-order products 
	$\hat{L}_j^\dag\hat{L}_k$ for $j,k=1,\ldots,R$, it was coined as the \emph{Hamiltonian-not-in-Lindblad span (HNLS)}
	condition denoted concisely as
	\begin{equation}
		\hat{H}\notin\mathrm{span}_{\mathbb{R}}
					 \left\{\openone,\hat{L}_j^{\mathrm{H}},\hat{L}_j^{\mathrm{AH}},(\hat{L}_j^\dag\hat{L}_k)^{\mathrm{H}},
							             (\hat{L}_j^\dag\hat{L}_k)^{\mathrm{AH}}\right\}
		\eqdef\mathbb{S}.
	\end{equation}
	Here $\hat{L}_j^\mathrm{H}\defeq(\hat{L}_j+\hat{L}_j^\dag)/2$ and $\hat{L}_j^\mathrm{AH}\defeq(\hat{L}_j-\hat{L}_j^\dag)/2i$ denote 
	the Hermitian and anti-Hermitian components of $\hat{L}_j$ respectively and the same definition to 
	$(\hat{L}_j^\dag\hat{L}_k)^{\mathrm{H}}$ and $(\hat{L}_j^\dag\hat{L}_k)^{\mathrm{AH}}$. $\mathbb{S}$ is thus an Euclidean 
	space. Further, the Markovian noise is called \emph{commuting} if the Hamiltonian $\hat{H}$ and Lindblad operators $\hat{L}_j$s are
	mutually commuting, namely $\lbrack{\hat{H},\hat{L}_j}\rbrack=\lbrack{\hat{L}_j,\hat{L}_k}\rbrack=0$. 
	In the following, we will illustrate for qubit systems that if HNLS holds, HL can be restored via QEC without ancilla, regardless whether
	the Markovian noise is commuting or not. 

	\emph{QEC quantum metrology without ancilla.}---Suppose we employ $N$ probes which evolve independently according to 
	(\ref{single_probe_selv}). The Hamiltonian for the first probe is 
	$\hat{H}^{(1)}\defeq\hat{H}\otimes\openone^{\otimes{N-1}}$ with the corresponding Lindblad operators being 
	$\hat{L}_k^{(1)}\defeq\hat{L}_k\otimes\openone^{\otimes{N-1}}$, and so forth.  The total Hamiltonian reads
	\begin{equation}
		\hat{H}_{\mathrm{tot}}\defeq\sum_{s=1}^S\hat{H}^{(n)}.
	\end{equation}
	A QEC code $\mathcal{C}$ can render the noisy quantum channel to effectively noiseless if the projection onto code space 
	$\hat{\Pi}_\mathrm{C}$ satisfies the QEC condition~\cite{Ekert1996,bennett1996mixed-state,Knill1997,nielsen2010quantum}
	\begin{equation} 
		\hat{\Pi}_\mathrm{C}\hat{L}_j^{(n)}\hat{\Pi}_\mathrm{C} = \lambda_k^{(n)}\hat{\Pi}_\mathrm{C} 
		\,\,\textrm{and}\,\,
		\hat{\Pi}_\mathrm{C}\lbrack\hat{L}_j^{(n)}\rbrack^\dag\hat{L}_k^{(m)}\hat{\Pi}_\mathrm{C} 
		= \mu_{jk}^{(nm)}\hat{\Pi}_\mathrm{C},
		\label{codeSpace_cond1+2}
	\end{equation}
	for every $n,m=1,\ldots,N$ and $j,k=1,\ldots,S$.
	The parameter-imprinting unitary channel can survive the QEC procedure $\mathcal{C}$ if
	\begin{equation} 
		\hat{\Pi}_\mathrm{C}\hat{H}_{\mathrm{tot}}\hat{\Pi}_\mathrm{C} \not\propto \hat{\Pi}_\mathrm{C}.
		\label{codeSpace_cond3}
	\end{equation}
	These conditions (\ref{codeSpace_cond1+2}) and (\ref{codeSpace_cond3}) are very similar to their counter part in the situation
	ancilla systems are plentiful~\cite{Sekatski2017a,Demkowicz-Dobrzanski2017,Zhou2018achieving}. And so is the proof which we outline 
	in the following. 
	
	Let us divide the total probing  
	time $T$ into $D$ pieces such that we carry out a QEC procedure via fast control after every $\delta{t}{\defeq}T/D$ evolution
	\begin{equation}
		\mathcal{E}_{\delta{t}}^{\otimes{N}}(\hat{\rho}_{\mathrm{tot}})
		\defeq \hat{K}_0\hat{\rho}_{\mathrm{tot}}\hat{K}_0^\dag 
		+ \sum_{k=1}^R\sum_{n=1}^N\hat{K}_k^{(n)}\hat{\rho}_{\mathrm{tot}}\lbrack\hat{K}_k^{(n)}\rbrack^\dag
	\end{equation}
	where
	\begin{equation}
		\hat{K}_0 \defeq \openone - \left(i\omega\hat{H}_{\mathrm{tot}}
										 +\frac{1}{2}\sum_{n=1}^N\sum_{k=1}^R\lbrack\hat{L}_k^{(n)}\rbrack^\dag\hat{L}_k^{(n)}\right)\delta{t}, 
	\end{equation} 
	and
	\begin{equation}
		\hat{K}_k^{(n)} \defeq \hat{L}_k^{(n)}\sqrt{\delta{t}}.  
	\end{equation}
	There exists a QEC code $\mathcal{C}$ capable of rendering $\mathcal{E}_{\delta{t}}^{\otimes{N}}$ to an effective unitary evolution generated by 
	the effective total Hamiltonian 
	$\hat{H}_\mathrm{eff}\defeq\hat{\Pi}_\mathrm{C}\hat{H}_{\mathrm{tot}}\hat{\Pi}_\mathrm{C}$
	\begin{equation}
		\mathcal{C}\circ\mathcal{E}_{\ud{t}}^{\otimes{N}}\left(\hat{\rho}_{\mathrm{tot}}\right)
		=\hat{\rho}_{\mathrm{tot}} - i\omega\lbrack{\hat{H}_\mathrm{eff},\hat{\rho}_{\mathrm{tot}}}\rbrack
		+\mathcal{O}(\delta{t}^2). 
	\end{equation}
	One can verify this by using the three conditions
	presented in (\ref{codeSpace_cond1+2}) and (\ref{codeSpace_cond3}) directly. Firstly, the diagnosis measurement would collapse the 
	probes onto either
	\begin{eqnarray}
		\hat{\Pi}_\mathrm{C}\mathcal{E}_{\delta{t}}(\hat{\rho}_{\mathrm{tot}})\hat{\Pi}_\mathrm{C}
		&=& \left(1+\sum_{n=1}^N\sum_{k=1}^R(|\lambda_k^{(n)}|^2-\mu_{kk}^{(nn)})\delta{t}\right)\hat{\rho}_{\mathrm{tot}} \nonumber\\
		&& - i\omega\lbrack{\hat{H}_\mathrm{eff},\hat{\rho}_{\mathrm{tot}}}\rbrack
		  +\mathcal{O}(\delta{t}^2)
	\end{eqnarray}
	if everything is all right, or 
	\begin{equation}
		\hat{\Pi}_\mathrm{E}\mathcal{E}_{\ud{t}}\left(\hat{\rho}_{\mathrm{tot}}\right)\hat{\Pi}_\mathrm{E} 
		=\sum_{n=1}^N\sum_{k=1}^R\hat{E}_k^{(n)}\hat{\rho}_{\mathrm{tot}}\lbrack{\hat{E}_k^{(n)}}\rbrack^\dag+\mathcal{O}(\delta{t}^2)
		\label{detectingError}
	\end{equation}
	with $\hat{E}_k^{(n)} \defeq \hat{\Pi}_\mathrm{E}\hat{L}_k^{(s)}\hat{\Pi}_\mathrm{C}\sqrt{\delta{t}}$ if error has been detected. 
	There exist a QEC code to correct error (\ref{detectingError}) according to the QEC 
	condition~\cite{Ekert1996,bennett1996mixed-state,Knill1997,nielsen2010quantum}
	\begin{equation}
		\hat{\Pi}_\mathrm{C}\lbrack\hat{E}_j^{(n)}\rbrack^\dag\hat{E}_k^{(m)}\hat{\Pi}_\mathrm{C}
		=\hat{\Pi}_\mathrm{C}\left(\mu^{(nm)}_{jk}-\lbrack\lambda_j^{(n)}\rbrack^*\lambda^{(m)}_k\right)\delta{t}.
	\end{equation}
\emph{Example.}--- Consider the Hamiltonian coupled to the parameter $\omega$ is 
	\begin{equation} 
		\hat{H}=\begin{pmatrix} 0 & 0 & i \\ 0 & 0 & 0 \\ -i & 0 & 0\end{pmatrix}.  
		\label{three_level_sys_Hamiltonian_cE}
	\end{equation} 
	The Markovian noise infesting the probe is described by 
	\begin{equation} 
		\hat{L}_1 = \begin{pmatrix} 0 & 1 & 0 \\ 1 & 0 & 0 \\ 0 & 0 & 0\end{pmatrix},\, 
		\hat{L}_2 = \begin{pmatrix} 0 & -i & 0 \\ i & 0 & 0 \\ 0 & 0 & 0\end{pmatrix}, \,\textrm{and}\, 
		\hat{L}_3 = \begin{pmatrix} 0 & 0 & 1 \\ 0 & 0 & 0 \\ 1 & 0 & 0\end{pmatrix}.  
		\label{Lindblad_operator_def} 
	\end{equation} 
	It has been condemned beyond saving by any ancilla-free QEC in the Supplementary Material of Ref.~\cite{Layden2019} when only 
	one probe is 
	available. The interesting part is that HNLS holds in this scenario. This condemnation can be avoid if at least $3$ probes are
	at our disposal.  One can verify that the three conditions presented in (\ref{codeSpace_cond1+2}) and (\ref{codeSpace_cond3}) 
	can be satisfied by the QEC code with logic states 
	\begin{equation*} 
		\ket{0_\mathrm{L}} = \ket{\circlearrowleft}^{\otimes{N}} 
		\quad\textrm{and}\quad
		\ket{1_\mathrm{L}} = \ket{\circlearrowright}^{\otimes{N}}.
	\end{equation*}
	Here $\ket{\circlearrowleft}\defeq\frac{1}{\sqrt{2}}\begin{pmatrix}1&0&i\end{pmatrix}^\mathrm{T}$
	and $\ket{\circlearrowright}\defeq\frac{1}{\sqrt{2}}\begin{pmatrix}1&0&-i\end{pmatrix}^\mathrm{T}$ are eigenvectors of 
	$\hat{H}$ belonging to eigenvalue $\pm1$ respectively. With $\ket{\Psi}=\frac{1}{\sqrt{2}}(\ket{0}_\mathrm{L}+\ket{1}_\mathrm{L})$ 
	fed, the optimal precision is 
	$\Delta{\omega}^2 = 1/4N^2T^2$ which is identical to the ultimate precision in the noiseless case~\cite{giovannetti2006quantum}.

\emph{Qubit probes.}---Suppose $\hat{H}_\perp$ is the component of $\hat{H}$ perpendicular to $\mathbb{S}$. Its spectrum decomposition 
	is 
	\begin{equation}
		\hat{H}_\perp =\|\hat{H}_\perp\|\left(\ket{\psi_+}\bra{\psi_+}-\ket{\psi_-}\bra{\psi_-}\right)
	\end{equation}
	with $\braket{\psi_\pm|\psi_\pm}=1$, $\braket{\psi_+|\psi_-}=0$ and $\|\bullet\|$ being the operator norm. An explicit QEC code that 
	can eradicate the error (\ref{detectingError}) is 
	\begin{equation}
		\hat{\Pi}_\mathrm{C}
		\defeq\left(\ket{\psi_+}\bra{\psi_+}\right)^{\otimes{N}}+\left(\ket{\psi_-}\bra{\psi_-}\right)^{\otimes{N}},
		\label{proj_CodeS_single-para_noAnc}
	\end{equation}
	when at least $3$ probes has been deployed. QEC condition (\ref{codeSpace_cond1+2}) can be ensured, since 
	$\braket{\psi_+|\hat{L}_j|\psi_+}=\braket{\psi_-|\hat{L}_j|\psi_-}$ and 
	$\braket{\psi_+|\hat{L}_j^\dag\hat{L}_k|\psi_+}=\braket{\psi_-|\hat{L}_j^\dag\hat{L}_k|\psi_-}$.
	And the effective total Hamiltonian after QEC would be 
	\begin{equation}
		\hat{H}_\mathrm{eff} = \sum_{n=1}^N\hat{H}_\perp^{(n)},
	\end{equation}
	which is non-vanishing as long as HNLS holds. The optimal input state is
	$\ket{\psi_\mathrm{opt}}=\frac{1}{\sqrt{2}}(\ket{\psi_+}^{\otimes{N}}+\ket{\psi_-}^{\otimes{N}})$ and the corresponding ultimate
	precision is 
	\begin{equation}
		\Delta^2\omega \ge {1}/{N^2T^2\|\hat{H}\|}.
	\end{equation}
	Thus HL is restored in both the sense of probing time $T$ as well as the number of probes.

%==================================================================================================================================
% Conclusion
%==================================================================================================================================
	\emph{Conclusion and discussion.}---When noiseless ancilla is inaccessible as in many situations, we show for qubit probes
	that HNLS is the necessary and sufficient condition for restoring HL via QEC. This is true for any Markovian noise whether 
	commuting or not. Our QEC code works for $N\ge3$. In stark contrast to the traditional QEC~\cite{nielsen2010quantum}, 
	the additional probing qubits are not merely redundant price paid for 
	fighting the detrimental Markovian noise. They would also do the parameter probing job at the same time. As a result HL
	is restored in the sense of both time consumed and number of qubits employed.
	We thus has partially resolved the open question raised previously in Ref.~\cite{Layden2019}. Our analysis
	also indicates that the parallel scheme can provide more possibility for QEC construction as can be seen from the example 
	presented in (\ref{three_level_sys_Hamiltonian_cE}) and (\ref{Lindblad_operator_def}).

%==================================================================================================================================
%	Acknowledge
%==================================================================================================================================
\begin{acknowledgments}
	This work was supported by Ministry of Science and Technology of China (Grants No. 2016YFA0302104 and 2016YFA0300600), National 
	Natural Science Foundation of China (Grant Nos. 91536108 and 11774406) and Chinese Academy of Sciences (XDPB08-3).
\end{acknowledgments}

\bibliography{Bibliography}

\begin{thebibliography}{53}%
\makeatletter
\providecommand \@ifxundefined [1]{%
 \@ifx{#1\undefined}
}%
\providecommand \@ifnum [1]{%
 \ifnum #1\expandafter \@firstoftwo
 \else \expandafter \@secondoftwo
 \fi
}%
\providecommand \@ifx [1]{%
 \ifx #1\expandafter \@firstoftwo
 \else \expandafter \@secondoftwo
 \fi
}%
\providecommand \natexlab [1]{#1}%
\providecommand \enquote  [1]{``#1''}%
\providecommand \bibnamefont  [1]{#1}%
\providecommand \bibfnamefont [1]{#1}%
\providecommand \citenamefont [1]{#1}%
\providecommand \href@noop [0]{\@secondoftwo}%
\providecommand \href [0]{\begingroup \@sanitize@url \@href}%
\providecommand \@href[1]{\@@startlink{#1}\@@href}%
\providecommand \@@href[1]{\endgroup#1\@@endlink}%
\providecommand \@sanitize@url [0]{\catcode `\\12\catcode `\$12\catcode
  `\&12\catcode `\#12\catcode `\^12\catcode `\_12\catcode `\%12\relax}%
\providecommand \@@startlink[1]{}%
\providecommand \@@endlink[0]{}%
\providecommand \url  [0]{\begingroup\@sanitize@url \@url }%
\providecommand \@url [1]{\endgroup\@href {#1}{\urlprefix }}%
\providecommand \urlprefix  [0]{URL }%
\providecommand \Eprint [0]{\href }%
\providecommand \doibase [0]{http://dx.doi.org/}%
\providecommand \selectlanguage [0]{\@gobble}%
\providecommand \bibinfo  [0]{\@secondoftwo}%
\providecommand \bibfield  [0]{\@secondoftwo}%
\providecommand \translation [1]{[#1]}%
\providecommand \BibitemOpen [0]{}%
\providecommand \bibitemStop [0]{}%
\providecommand \bibitemNoStop [0]{.\EOS\space}%
\providecommand \EOS [0]{\spacefactor3000\relax}%
\providecommand \BibitemShut  [1]{\csname bibitem#1\endcsname}%
\let\auto@bib@innerbib\@empty
%</preamble>
\bibitem [{\citenamefont {Grumbling}\ and\ \citenamefont
  {Horowitz}(2018)}]{usa2018NAS}%
  \BibitemOpen
  \bibinfo {editor} {\bibfnamefont {E.}~\bibnamefont {Grumbling}}\ and\
  \bibinfo {editor} {\bibfnamefont {M.}~\bibnamefont {Horowitz}},\ eds.,\ \href
  {\doibase 10.17226/25196} {\emph {\bibinfo {title} {National Academies of
  Sciences, Engineering, and Medicine. Quantum Computing: Progress and
  Prospects}}}\ (\bibinfo  {publisher} {The National Academies Press},\
  \bibinfo {address} {Washington, DC},\ \bibinfo {year} {2018})\BibitemShut
  {NoStop}%
\bibitem [{\citenamefont {Giovannetti}\ \emph {et~al.}(2004)\citenamefont
  {Giovannetti}, \citenamefont {Lloyd},\ and\ \citenamefont
  {Maccone}}]{giovannetti2004quantum-enhanced}%
  \BibitemOpen
  \bibfield  {author} {\bibinfo {author} {\bibfnamefont {V.}~\bibnamefont
  {Giovannetti}}, \bibinfo {author} {\bibfnamefont {S.}~\bibnamefont {Lloyd}},
  \ and\ \bibinfo {author} {\bibfnamefont {L.}~\bibnamefont {Maccone}},\ }\href
  {\doibase 10.1126/science.1104149} {\bibfield  {journal} {\bibinfo  {journal}
  {Science}\ }\textbf {\bibinfo {volume} {306}},\ \bibinfo {pages} {1330}
  (\bibinfo {year} {2004})}\BibitemShut {NoStop}%
\bibitem [{\citenamefont {Giovannetti}\ \emph {et~al.}(2006)\citenamefont
  {Giovannetti}, \citenamefont {Lloyd},\ and\ \citenamefont
  {Maccone}}]{giovannetti2006quantum}%
  \BibitemOpen
  \bibfield  {author} {\bibinfo {author} {\bibfnamefont {V.}~\bibnamefont
  {Giovannetti}}, \bibinfo {author} {\bibfnamefont {S.}~\bibnamefont {Lloyd}},
  \ and\ \bibinfo {author} {\bibfnamefont {L.}~\bibnamefont {Maccone}},\ }\href
  {\doibase 10.1103/PhysRevLett.96.010401} {\bibfield  {journal} {\bibinfo
  {journal} {Phys. Rev. Lett.}\ }\textbf {\bibinfo {volume} {96}},\ \bibinfo
  {pages} {010401} (\bibinfo {year} {2006})}\BibitemShut {NoStop}%
\bibitem [{\citenamefont {Demkowicz-Dobrza{\'{n}}ski}\ \emph
  {et~al.}(2015)\citenamefont {Demkowicz-Dobrza{\'{n}}ski}, \citenamefont
  {Jarzyna},\ and\ \citenamefont
  {Ko{\l}ody{\'{n}}ski}}]{Demkowicz-Dobrzanski2015}%
  \BibitemOpen
  \bibfield  {author} {\bibinfo {author} {\bibfnamefont {R.}~\bibnamefont
  {Demkowicz-Dobrza{\'{n}}ski}}, \bibinfo {author} {\bibfnamefont
  {M.}~\bibnamefont {Jarzyna}}, \ and\ \bibinfo {author} {\bibfnamefont
  {J.}~\bibnamefont {Ko{\l}ody{\'{n}}ski}},\ }in\ \href {\doibase
  10.1016/bs.po.2015.02.003} {\emph {\bibinfo {booktitle} {Prog. Opt.}}},\
  Vol.~\bibinfo {volume} {60}\ (\bibinfo  {publisher} {Elsevier},\ \bibinfo
  {year} {2015})\ pp.\ \bibinfo {pages} {345--435}\BibitemShut {NoStop}%
\bibitem [{\citenamefont {Degen}\ \emph {et~al.}(2017)\citenamefont {Degen},
  \citenamefont {Reinhard},\ and\ \citenamefont {Cappellaro}}]{Degen2017}%
  \BibitemOpen
  \bibfield  {author} {\bibinfo {author} {\bibfnamefont {C.~L.}\ \bibnamefont
  {Degen}}, \bibinfo {author} {\bibfnamefont {F.}~\bibnamefont {Reinhard}}, \
  and\ \bibinfo {author} {\bibfnamefont {P.}~\bibnamefont {Cappellaro}},\
  }\href {\doibase 10.1103/RevModPhys.89.035002} {\bibfield  {journal}
  {\bibinfo  {journal} {Rev. Mod. Phys.}\ }\textbf {\bibinfo {volume} {89}},\
  \bibinfo {pages} {035002} (\bibinfo {year} {2017})}\BibitemShut {NoStop}%
\bibitem [{\citenamefont {Pezz{\`{e}}}\ \emph {et~al.}(2018)\citenamefont
  {Pezz{\`{e}}}, \citenamefont {Smerzi}, \citenamefont {Oberthaler},
  \citenamefont {Schmied},\ and\ \citenamefont {Treutlein}}]{Pezze2018}%
  \BibitemOpen
  \bibfield  {author} {\bibinfo {author} {\bibfnamefont {L.}~\bibnamefont
  {Pezz{\`{e}}}}, \bibinfo {author} {\bibfnamefont {A.}~\bibnamefont {Smerzi}},
  \bibinfo {author} {\bibfnamefont {M.~K.}\ \bibnamefont {Oberthaler}},
  \bibinfo {author} {\bibfnamefont {R.}~\bibnamefont {Schmied}}, \ and\
  \bibinfo {author} {\bibfnamefont {P.}~\bibnamefont {Treutlein}},\ }\href
  {\doibase 10.1103/RevModPhys.90.035005} {\bibfield  {journal} {\bibinfo
  {journal} {Rev. Mod. Phys.}\ }\textbf {\bibinfo {volume} {90}},\ \bibinfo
  {pages} {035005} (\bibinfo {year} {2018})}\BibitemShut {NoStop}%
\bibitem [{\citenamefont {Kwon}\ \emph {et~al.}(2019)\citenamefont {Kwon},
  \citenamefont {Tan}, \citenamefont {Volkoff},\ and\ \citenamefont
  {Jeong}}]{Kwon2019}%
  \BibitemOpen
  \bibfield  {author} {\bibinfo {author} {\bibfnamefont {H.}~\bibnamefont
  {Kwon}}, \bibinfo {author} {\bibfnamefont {K.~C.}\ \bibnamefont {Tan}},
  \bibinfo {author} {\bibfnamefont {T.}~\bibnamefont {Volkoff}}, \ and\
  \bibinfo {author} {\bibfnamefont {H.}~\bibnamefont {Jeong}},\ }\href
  {\doibase 10.1103/PhysRevLett.122.040503} {\bibfield  {journal} {\bibinfo
  {journal} {Phys. Rev. Lett.}\ }\textbf {\bibinfo {volume} {122}},\ \bibinfo
  {pages} {040503} (\bibinfo {year} {2019})}\BibitemShut {NoStop}%
\bibitem [{\citenamefont {Bollinger}\ \emph {et~al.}(1999)\citenamefont
  {Bollinger}, \citenamefont {Itano}, \citenamefont {Wineland},\ and\
  \citenamefont {Heinzen}}]{Bollinger1996}%
  \BibitemOpen
  \bibfield  {author} {\bibinfo {author} {\bibfnamefont {J.~J.}\ \bibnamefont
  {Bollinger}}, \bibinfo {author} {\bibfnamefont {W.~M.}\ \bibnamefont
  {Itano}}, \bibinfo {author} {\bibfnamefont {D.~J.}\ \bibnamefont {Wineland}},
  \ and\ \bibinfo {author} {\bibfnamefont {D.~J.}\ \bibnamefont {Heinzen}},\
  }\href {\doibase 10.1103/PhysRevA.54.R4649} {\bibfield  {journal} {\bibinfo
  {journal} {Phys. Rev. A}\ }\textbf {\bibinfo {volume} {54}},\ \bibinfo
  {pages} {R4649} (\bibinfo {year} {1999})}\BibitemShut {NoStop}%
\bibitem [{\citenamefont {Kessler}\ \emph {et~al.}(2014)\citenamefont
  {Kessler}, \citenamefont {K{\'{o}}m{\'{a}}r}, \citenamefont {Bishof},
  \citenamefont {Jiang}, \citenamefont {S{\o}rensen}, \citenamefont {Ye},\ and\
  \citenamefont {Lukin}}]{Kessler2014}%
  \BibitemOpen
  \bibfield  {author} {\bibinfo {author} {\bibfnamefont {E.~M.}\ \bibnamefont
  {Kessler}}, \bibinfo {author} {\bibfnamefont {P.}~\bibnamefont
  {K{\'{o}}m{\'{a}}r}}, \bibinfo {author} {\bibfnamefont {M.}~\bibnamefont
  {Bishof}}, \bibinfo {author} {\bibfnamefont {L.}~\bibnamefont {Jiang}},
  \bibinfo {author} {\bibfnamefont {A.~S.}\ \bibnamefont {S{\o}rensen}},
  \bibinfo {author} {\bibfnamefont {J.}~\bibnamefont {Ye}}, \ and\ \bibinfo
  {author} {\bibfnamefont {M.~D.}\ \bibnamefont {Lukin}},\ }\href {\doibase
  10.1103/PhysRevLett.112.190403} {\bibfield  {journal} {\bibinfo  {journal}
  {Phys. Rev. Lett.}\ }\textbf {\bibinfo {volume} {112}},\ \bibinfo {pages}
  {190403} (\bibinfo {year} {2014})}\BibitemShut {NoStop}%
\bibitem [{\citenamefont {Derevianko}\ and\ \citenamefont
  {Katori}(2011)}]{Derevianko2011}%
  \BibitemOpen
  \bibfield  {author} {\bibinfo {author} {\bibfnamefont {A.}~\bibnamefont
  {Derevianko}}\ and\ \bibinfo {author} {\bibfnamefont {H.}~\bibnamefont
  {Katori}},\ }\href {\doibase 10.1103/RevModPhys.83.331} {\bibfield  {journal}
  {\bibinfo  {journal} {Rev. Mod. Phys.}\ }\textbf {\bibinfo {volume} {83}},\
  \bibinfo {pages} {331} (\bibinfo {year} {2011})}\BibitemShut {NoStop}%
\bibitem [{\citenamefont {Zhang}\ and\ \citenamefont {Ye}(2016)}]{Zhang2016}%
  \BibitemOpen
  \bibfield  {author} {\bibinfo {author} {\bibfnamefont {X.}~\bibnamefont
  {Zhang}}\ and\ \bibinfo {author} {\bibfnamefont {J.}~\bibnamefont {Ye}},\
  }\href {\doibase 10.1093/nsr/nww013} {\bibfield  {journal} {\bibinfo
  {journal} {Natl. Sci. Rev.}\ }\textbf {\bibinfo {volume} {3}},\ \bibinfo
  {pages} {189} (\bibinfo {year} {2016})}\BibitemShut {NoStop}%
\bibitem [{\citenamefont {Collaboration}(2013)}]{LIGO2013}%
  \BibitemOpen
  \bibfield  {author} {\bibinfo {author} {\bibfnamefont {The LIGO Scientific}\
  \bibnamefont {Collaboration}},\ }\href {\doibase 10.1038/nphoton.2013.177}
  {\bibfield  {journal} {\bibinfo  {journal} {Nat. Photonics}\ }\textbf
  {\bibinfo {volume} {7}},\ \bibinfo {pages} {613} (\bibinfo {year}
  {2013})}\BibitemShut {NoStop}%
\bibitem [{\citenamefont {Schnabel}\ \emph {et~al.}(2010)\citenamefont
  {Schnabel}, \citenamefont {Mavalvala}, \citenamefont {McClelland},\ and\
  \citenamefont {Lam}}]{Schnabel2010}%
  \BibitemOpen
  \bibfield  {author} {\bibinfo {author} {\bibfnamefont {R.}~\bibnamefont
  {Schnabel}}, \bibinfo {author} {\bibfnamefont {N.}~\bibnamefont {Mavalvala}},
  \bibinfo {author} {\bibfnamefont {D.~E.}\ \bibnamefont {McClelland}}, \ and\
  \bibinfo {author} {\bibfnamefont {P.~K.}\ \bibnamefont {Lam}},\ }\href
  {\doibase 10.1038/ncomms1122} {\bibfield  {journal} {\bibinfo  {journal}
  {Nat. Commun.}\ }\textbf {\bibinfo {volume} {1}},\ \bibinfo {pages} {121}
  (\bibinfo {year} {2010})}\BibitemShut {NoStop}%
\bibitem [{\citenamefont {Taylor}\ and\ \citenamefont
  {Bowen}(2016)}]{Taylor2016}%
  \BibitemOpen
  \bibfield  {author} {\bibinfo {author} {\bibfnamefont {M.~A.}\ \bibnamefont
  {Taylor}}\ and\ \bibinfo {author} {\bibfnamefont {W.~P.}\ \bibnamefont
  {Bowen}},\ }\href {\doibase 10.1016/j.physrep.2015.12.002} {\bibfield
  {journal} {\bibinfo  {journal} {Phys. Rep.}\ }\textbf {\bibinfo {volume}
  {615}},\ \bibinfo {pages} {1} (\bibinfo {year} {2016})}\BibitemShut {NoStop}%
\bibitem [{\citenamefont {Huelga}\ \emph {et~al.}(1997)\citenamefont {Huelga},
  \citenamefont {Macchiavello}, \citenamefont {Pellizzari}, \citenamefont
  {Ekert}, \citenamefont {Plenio},\ and\ \citenamefont {Cirac}}]{Huelga1997}%
  \BibitemOpen
  \bibfield  {author} {\bibinfo {author} {\bibfnamefont {S.~F.}\ \bibnamefont
  {Huelga}}, \bibinfo {author} {\bibfnamefont {C.}~\bibnamefont
  {Macchiavello}}, \bibinfo {author} {\bibfnamefont {T.}~\bibnamefont
  {Pellizzari}}, \bibinfo {author} {\bibfnamefont {A.~K.}\ \bibnamefont
  {Ekert}}, \bibinfo {author} {\bibfnamefont {M.~B.}\ \bibnamefont {Plenio}}, \
  and\ \bibinfo {author} {\bibfnamefont {J.~I.}\ \bibnamefont {Cirac}},\ }\href
  {\doibase 10.1103/PhysRevLett.79.3865} {\bibfield  {journal} {\bibinfo
  {journal} {Phys. Rev. Lett.}\ }\textbf {\bibinfo {volume} {79}},\ \bibinfo
  {pages} {3865} (\bibinfo {year} {1997})}\BibitemShut {NoStop}%
\bibitem [{\citenamefont {Escher}\ \emph {et~al.}(2011)\citenamefont {Escher},
  \citenamefont {{de Matos Filho}},\ and\ \citenamefont
  {Davidovich}}]{Escher2011general}%
  \BibitemOpen
  \bibfield  {author} {\bibinfo {author} {\bibfnamefont {B.~M.}\ \bibnamefont
  {Escher}}, \bibinfo {author} {\bibfnamefont {R.~L.}\ \bibnamefont {{de Matos
  Filho}}}, \ and\ \bibinfo {author} {\bibfnamefont {L.}~\bibnamefont
  {Davidovich}},\ }\href {\doibase 10.1038/nphys1958} {\bibfield  {journal}
  {\bibinfo  {journal} {Nat. Phys.}\ }\textbf {\bibinfo {volume} {7}},\
  \bibinfo {pages} {406} (\bibinfo {year} {2011})}\BibitemShut {NoStop}%
\bibitem [{\citenamefont {Chin}\ \emph {et~al.}(2012)\citenamefont {Chin},
  \citenamefont {Huelga},\ and\ \citenamefont {Plenio}}]{Chin2012}%
  \BibitemOpen
  \bibfield  {author} {\bibinfo {author} {\bibfnamefont {A.~W.}\ \bibnamefont
  {Chin}}, \bibinfo {author} {\bibfnamefont {S.~F.}\ \bibnamefont {Huelga}}, \
  and\ \bibinfo {author} {\bibfnamefont {M.~B.}\ \bibnamefont {Plenio}},\
  }\href {\doibase 10.1103/PhysRevLett.109.233601} {\bibfield  {journal}
  {\bibinfo  {journal} {Phys. Rev. Lett.}\ }\textbf {\bibinfo {volume} {109}},\
  \bibinfo {pages} {233601} (\bibinfo {year} {2012})}\BibitemShut {NoStop}%
\bibitem [{\citenamefont {Demkowicz-Dobrza{\'{n}}ski}\ \emph
  {et~al.}(2012)\citenamefont {Demkowicz-Dobrza{\'{n}}ski}, \citenamefont
  {Ko{\l}ody{\'{n}}ski},\ and\ \citenamefont {Gu{\c t}{\v
  a}}}]{Demkowicz-Dobrzanski2012}%
  \BibitemOpen
  \bibfield  {author} {\bibinfo {author} {\bibfnamefont {R.}~\bibnamefont
  {Demkowicz-Dobrza{\'{n}}ski}}, \bibinfo {author} {\bibfnamefont
  {J.}~\bibnamefont {Ko{\l}ody{\'{n}}ski}}, \ and\ \bibinfo {author}
  {\bibfnamefont {M.}~\bibnamefont {Gu{\c t}{\v a}}},\ }\href {\doibase
  10.1038/ncomms2067} {\bibfield  {journal} {\bibinfo  {journal} {Nat.
  Commun.}\ }\textbf {\bibinfo {volume} {3}},\ \bibinfo {pages} {1063}
  (\bibinfo {year} {2012})}\BibitemShut {NoStop}%
\bibitem [{\citenamefont {S{\o}rensen}\ \emph {et~al.}(2001)\citenamefont
  {S{\o}rensen}, \citenamefont {Duan}, \citenamefont {Cirac},\ and\
  \citenamefont {Zoller}}]{sorensen2001many}%
  \BibitemOpen
  \bibfield  {author} {\bibinfo {author} {\bibfnamefont {A.}~\bibnamefont
  {S{\o}rensen}}, \bibinfo {author} {\bibfnamefont {L.-M.}\ \bibnamefont
  {Duan}}, \bibinfo {author} {\bibfnamefont {J.~I.}\ \bibnamefont {Cirac}}, \
  and\ \bibinfo {author} {\bibfnamefont {P.}~\bibnamefont {Zoller}},\ }\href
  {\doibase 10.1038/35051038} {\bibfield  {journal} {\bibinfo  {journal}
  {Nature}\ }\textbf {\bibinfo {volume} {409}},\ \bibinfo {pages} {63}
  (\bibinfo {year} {2001})}\BibitemShut {NoStop}%
\bibitem [{\citenamefont {Dunningham}\ \emph {et~al.}(2002)\citenamefont
  {Dunningham}, \citenamefont {Burnett},\ and\ \citenamefont
  {Barnett}}]{Dunningham2002}%
  \BibitemOpen
  \bibfield  {author} {\bibinfo {author} {\bibfnamefont {J.~A.}\ \bibnamefont
  {Dunningham}}, \bibinfo {author} {\bibfnamefont {K.}~\bibnamefont {Burnett}},
  \ and\ \bibinfo {author} {\bibfnamefont {S.~M.}\ \bibnamefont {Barnett}},\
  }\href {\doibase 10.1103/PhysRevLett.89.150401} {\bibfield  {journal}
  {\bibinfo  {journal} {Phys. Rev. Lett.}\ }\textbf {\bibinfo {volume} {89}},\
  \bibinfo {pages} {150401} (\bibinfo {year} {2002})}\BibitemShut {NoStop}%
\bibitem [{\citenamefont {Ma}\ \emph {et~al.}(2011)\citenamefont {Ma},
  \citenamefont {Wang}, \citenamefont {Sun},\ and\ \citenamefont
  {Nori}}]{ma2011quantum}%
  \BibitemOpen
  \bibfield  {author} {\bibinfo {author} {\bibfnamefont {J.}~\bibnamefont
  {Ma}}, \bibinfo {author} {\bibfnamefont {X.}~\bibnamefont {Wang}}, \bibinfo
  {author} {\bibfnamefont {C.}~\bibnamefont {Sun}}, \ and\ \bibinfo {author}
  {\bibfnamefont {F.}~\bibnamefont {Nori}},\ }\href {\doibase
  10.1016/j.physrep.2011.08.003} {\bibfield  {journal} {\bibinfo  {journal}
  {Phys. Rep.}\ }\textbf {\bibinfo {volume} {509}},\ \bibinfo {pages} {89}
  (\bibinfo {year} {2011})}\BibitemShut {NoStop}%
\bibitem [{\citenamefont {Duan}(2011)}]{Duan2011}%
  \BibitemOpen
  \bibfield  {author} {\bibinfo {author} {\bibfnamefont {L.-M.}\ \bibnamefont
  {Duan}},\ }\href {\doibase 10.1103/PhysRevLett.107.180502} {\bibfield
  {journal} {\bibinfo  {journal} {Phys. Rev. Lett.}\ }\textbf {\bibinfo
  {volume} {107}},\ \bibinfo {pages} {180502} (\bibinfo {year}
  {2011})}\BibitemShut {NoStop}%
\bibitem [{\citenamefont {Zhang}\ and\ \citenamefont {Duan}(2013)}]{Zhang2013}%
  \BibitemOpen
  \bibfield  {author} {\bibinfo {author} {\bibfnamefont {Z.}~\bibnamefont
  {Zhang}}\ and\ \bibinfo {author} {\bibfnamefont {L.-M.}\ \bibnamefont
  {Duan}},\ }\href {\doibase 10.1103/PhysRevLett.111.180401} {\bibfield
  {journal} {\bibinfo  {journal} {Phys. Rev. Lett.}\ }\textbf {\bibinfo
  {volume} {111}},\ \bibinfo {pages} {180401} (\bibinfo {year}
  {2013})}\BibitemShut {NoStop}%
\bibitem [{\citenamefont {Pezz{\'{e}}}\ and\ \citenamefont
  {Smerzi}(2013)}]{Pezze2013}%
  \BibitemOpen
  \bibfield  {author} {\bibinfo {author} {\bibfnamefont {L.}~\bibnamefont
  {Pezz{\'{e}}}}\ and\ \bibinfo {author} {\bibfnamefont {A.}~\bibnamefont
  {Smerzi}},\ }\href {\doibase 10.1103/PhysRevLett.110.163604} {\bibfield
  {journal} {\bibinfo  {journal} {Phys. Rev. Lett.}\ }\textbf {\bibinfo
  {volume} {110}},\ \bibinfo {pages} {163604} (\bibinfo {year}
  {2013})}\BibitemShut {NoStop}%
\bibitem [{\citenamefont {Demkowicz-Dobrza{\'{n}}ski}\ and\ \citenamefont
  {Maccone}(2014)}]{Demkowicz-Dobrzanski2014}%
  \BibitemOpen
  \bibfield  {author} {\bibinfo {author} {\bibfnamefont {R.}~\bibnamefont
  {Demkowicz-Dobrza{\'{n}}ski}}\ and\ \bibinfo {author} {\bibfnamefont
  {L.}~\bibnamefont {Maccone}},\ }\href {\doibase
  10.1103/PhysRevLett.113.250801} {\bibfield  {journal} {\bibinfo  {journal}
  {Phys. Rev. Lett.}\ }\textbf {\bibinfo {volume} {113}},\ \bibinfo {pages}
  {250801} (\bibinfo {year} {2014})}\BibitemShut {NoStop}%
\bibitem [{\citenamefont {Chaves}\ \emph {et~al.}(2013)\citenamefont {Chaves},
  \citenamefont {Brask}, \citenamefont {Markiewicz}, \citenamefont
  {Ko{\l}ody{\'{n}}ski},\ and\ \citenamefont {Ac{\'{i}}n}}]{Chaves2013}%
  \BibitemOpen
  \bibfield  {author} {\bibinfo {author} {\bibfnamefont {R.}~\bibnamefont
  {Chaves}}, \bibinfo {author} {\bibfnamefont {J.~B.}\ \bibnamefont {Brask}},
  \bibinfo {author} {\bibfnamefont {M.}~\bibnamefont {Markiewicz}}, \bibinfo
  {author} {\bibfnamefont {J.}~\bibnamefont {Ko{\l}ody{\'{n}}ski}}, \ and\
  \bibinfo {author} {\bibfnamefont {A.}~\bibnamefont {Ac{\'{i}}n}},\ }\href
  {\doibase 10.1103/PhysRevLett.111.120401} {\bibfield  {journal} {\bibinfo
  {journal} {Phys. Rev. Lett.}\ }\textbf {\bibinfo {volume} {111}},\ \bibinfo
  {pages} {120401} (\bibinfo {year} {2013})}\BibitemShut {NoStop}%
\bibitem [{\citenamefont {Brask}\ \emph {et~al.}(2015)\citenamefont {Brask},
  \citenamefont {Chaves},\ and\ \citenamefont
  {Ko{\l}ody{\'{n}}ski}}]{Brask2015}%
  \BibitemOpen
  \bibfield  {author} {\bibinfo {author} {\bibfnamefont {J.~B.}\ \bibnamefont
  {Brask}}, \bibinfo {author} {\bibfnamefont {R.}~\bibnamefont {Chaves}}, \
  and\ \bibinfo {author} {\bibfnamefont {J.}~\bibnamefont
  {Ko{\l}ody{\'{n}}ski}},\ }\href {\doibase 10.1103/PhysRevX.5.031010}
  {\bibfield  {journal} {\bibinfo  {journal} {Phys. Rev. X}\ }\textbf {\bibinfo
  {volume} {5}},\ \bibinfo {pages} {031010} (\bibinfo {year}
  {2015})}\BibitemShut {NoStop}%
\bibitem [{\citenamefont {Hentschel}\ and\ \citenamefont
  {Sanders}(2010)}]{Hentschel2010}%
  \BibitemOpen
  \bibfield  {author} {\bibinfo {author} {\bibfnamefont {A.}~\bibnamefont
  {Hentschel}}\ and\ \bibinfo {author} {\bibfnamefont {B.~C.}\ \bibnamefont
  {Sanders}},\ }\href {\doibase 10.1103/PhysRevLett.104.063603} {\bibfield
  {journal} {\bibinfo  {journal} {Phys. Rev. Lett.}\ }\textbf {\bibinfo
  {volume} {104}},\ \bibinfo {pages} {063603} (\bibinfo {year}
  {2010})}\BibitemShut {NoStop}%
\bibitem [{\citenamefont {Hentschel}\ and\ \citenamefont
  {Sanders}(2011)}]{Hentschel2011}%
  \BibitemOpen
  \bibfield  {author} {\bibinfo {author} {\bibfnamefont {A.}~\bibnamefont
  {Hentschel}}\ and\ \bibinfo {author} {\bibfnamefont {B.~C.}\ \bibnamefont
  {Sanders}},\ }\href {\doibase 10.1103/PhysRevLett.107.233601} {\bibfield
  {journal} {\bibinfo  {journal} {Phys. Rev. Lett.}\ }\textbf {\bibinfo
  {volume} {107}},\ \bibinfo {pages} {233601} (\bibinfo {year}
  {2011})}\BibitemShut {NoStop}%
\bibitem [{\citenamefont {Lovett}\ \emph {et~al.}(2013)\citenamefont {Lovett},
  \citenamefont {Crosnier}, \citenamefont {Perarnau-Llobet},\ and\
  \citenamefont {Sanders}}]{Lovett2013}%
  \BibitemOpen
  \bibfield  {author} {\bibinfo {author} {\bibfnamefont {N.~B.}\ \bibnamefont
  {Lovett}}, \bibinfo {author} {\bibfnamefont {C.}~\bibnamefont {Crosnier}},
  \bibinfo {author} {\bibfnamefont {M.}~\bibnamefont {Perarnau-Llobet}}, \ and\
  \bibinfo {author} {\bibfnamefont {B.~C.}\ \bibnamefont {Sanders}},\ }\href
  {\doibase 10.1103/PhysRevLett.110.220501} {\bibfield  {journal} {\bibinfo
  {journal} {Phys. Rev. Lett.}\ }\textbf {\bibinfo {volume} {110}},\ \bibinfo
  {pages} {220501} (\bibinfo {year} {2013})}\BibitemShut {NoStop}%
\bibitem [{\citenamefont {Palittapongarnpim}\ \emph {et~al.}(2017)\citenamefont
  {Palittapongarnpim}, \citenamefont {Wittek}, \citenamefont {Zahedinejad},
  \citenamefont {Vedaie},\ and\ \citenamefont
  {Sanders}}]{Palittapongarnpim2017}%
  \BibitemOpen
  \bibfield  {author} {\bibinfo {author} {\bibfnamefont {P.}~\bibnamefont
  {Palittapongarnpim}}, \bibinfo {author} {\bibfnamefont {P.}~\bibnamefont
  {Wittek}}, \bibinfo {author} {\bibfnamefont {E.}~\bibnamefont {Zahedinejad}},
  \bibinfo {author} {\bibfnamefont {S.}~\bibnamefont {Vedaie}}, \ and\ \bibinfo
  {author} {\bibfnamefont {B.~C.}\ \bibnamefont {Sanders}},\ }\href {\doibase
  10.1016/j.neucom.2016.12.087} {\bibfield  {journal} {\bibinfo  {journal}
  {Neurocomputing}\ }\textbf {\bibinfo {volume} {268}},\ \bibinfo {pages} {116}
  (\bibinfo {year} {2017})}\BibitemShut {NoStop}%
\bibitem [{\citenamefont {Lumino}\ \emph {et~al.}(2018)\citenamefont {Lumino},
  \citenamefont {Polino}, \citenamefont {Rab}, \citenamefont {Spagnolo},
  \citenamefont {Wiebe},\ and\ \citenamefont {Sciarrino}}]{Lumino2018}%
  \BibitemOpen
  \bibfield  {author} {\bibinfo {author} {\bibfnamefont {A.}~\bibnamefont
  {Lumino}}, \bibinfo {author} {\bibfnamefont {E.}~\bibnamefont {Polino}},
  \bibinfo {author} {\bibfnamefont {G.}~\bibnamefont {Rab}, \bibfnamefont {Adil
  S.~andMilani}}, \bibinfo {author} {\bibfnamefont {N.}~\bibnamefont
  {Spagnolo}}, \bibinfo {author} {\bibfnamefont {N.}~\bibnamefont {Wiebe}}, \
  and\ \bibinfo {author} {\bibfnamefont {F.}~\bibnamefont {Sciarrino}},\ }\href
  {\doibase 10.1103/PhysRevApplied.10.044033} {\bibfield  {journal} {\bibinfo
  {journal} {Phys. Rev. Applied}\ }\textbf {\bibinfo {volume} {10}},\ \bibinfo
  {pages} {044033} (\bibinfo {year} {2018})}\BibitemShut {NoStop}%
\bibitem [{\citenamefont {Viola}\ \emph
  {et~al.}(1999{\natexlab{a}})\citenamefont {Viola}, \citenamefont {Knill},\
  and\ \citenamefont {Lloyd}}]{Viola1999a}%
  \BibitemOpen
  \bibfield  {author} {\bibinfo {author} {\bibfnamefont {L.}~\bibnamefont
  {Viola}}, \bibinfo {author} {\bibfnamefont {E.}~\bibnamefont {Knill}}, \ and\
  \bibinfo {author} {\bibfnamefont {S.}~\bibnamefont {Lloyd}},\ }\href
  {\doibase 10.1103/PhysRevLett.82.2417} {\bibfield  {journal} {\bibinfo
  {journal} {Phys. Rev. Lett.}\ }\textbf {\bibinfo {volume} {82}},\ \bibinfo
  {pages} {2417} (\bibinfo {year} {1999}{\natexlab{a}})}\BibitemShut {NoStop}%
\bibitem [{\citenamefont {Viola}\ \emph
  {et~al.}(1999{\natexlab{b}})\citenamefont {Viola}, \citenamefont {Lloyd},\
  and\ \citenamefont {Knill}}]{Viola1999b}%
  \BibitemOpen
  \bibfield  {author} {\bibinfo {author} {\bibfnamefont {L.}~\bibnamefont
  {Viola}}, \bibinfo {author} {\bibfnamefont {S.}~\bibnamefont {Lloyd}}, \ and\
  \bibinfo {author} {\bibfnamefont {E.}~\bibnamefont {Knill}},\ }\href
  {\doibase 10.1103/PhysRevLett.83.4888} {\bibfield  {journal} {\bibinfo
  {journal} {Phys. Rev. Lett.}\ }\textbf {\bibinfo {volume} {83}},\ \bibinfo
  {pages} {4888} (\bibinfo {year} {1999}{\natexlab{b}})}\BibitemShut {NoStop}%
\bibitem [{\citenamefont {Tan}\ \emph {et~al.}(2013)\citenamefont {Tan},
  \citenamefont {Huang}, \citenamefont {Yin}, \citenamefont {Kuang},\ and\
  \citenamefont {Wang}}]{Tan2013}%
  \BibitemOpen
  \bibfield  {author} {\bibinfo {author} {\bibfnamefont {Q.-S.}\ \bibnamefont
  {Tan}}, \bibinfo {author} {\bibfnamefont {Y.}~\bibnamefont {Huang}}, \bibinfo
  {author} {\bibfnamefont {X.}~\bibnamefont {Yin}}, \bibinfo {author}
  {\bibfnamefont {L.-M.}\ \bibnamefont {Kuang}}, \ and\ \bibinfo {author}
  {\bibfnamefont {X.}~\bibnamefont {Wang}},\ }\href {\doibase
  10.1103/PhysRevA.87.032102} {\bibfield  {journal} {\bibinfo  {journal} {Phys.
  Rev. A}\ }\textbf {\bibinfo {volume} {87}},\ \bibinfo {pages} {032102}
  (\bibinfo {year} {2013})}\BibitemShut {NoStop}%
\bibitem [{\citenamefont {Sekatski}\ \emph {et~al.}(2016)\citenamefont
  {Sekatski}, \citenamefont {Skotiniotis},\ and\ \citenamefont
  {D{\"{u}}r}}]{Sekatski2016}%
  \BibitemOpen
  \bibfield  {author} {\bibinfo {author} {\bibfnamefont {P.}~\bibnamefont
  {Sekatski}}, \bibinfo {author} {\bibfnamefont {M.}~\bibnamefont
  {Skotiniotis}}, \ and\ \bibinfo {author} {\bibfnamefont {W.}~\bibnamefont
  {D{\"{u}}r}},\ }\href {\doibase 10.1088/1367-2630/18/7/073034} {\bibfield
  {journal} {\bibinfo  {journal} {New J. Phys.}\ }\textbf {\bibinfo {volume}
  {18}},\ \bibinfo {pages} {073034} (\bibinfo {year} {2016})}\BibitemShut
  {NoStop}%
\bibitem [{\citenamefont {Unden}\ \emph {et~al.}(2016)\citenamefont {Unden},
  \citenamefont {Balasubramanian}, \citenamefont {Louzon}, \citenamefont
  {Vinkler}, \citenamefont {Plenio}, \citenamefont {Markham}, \citenamefont
  {Twitchen}, \citenamefont {Stacey}, \citenamefont {Lovchinsky}, \citenamefont
  {Sushkov}, \citenamefont {Lukin}, \citenamefont {Retzker}, \citenamefont
  {Naydenov}, \citenamefont {McGuinness},\ and\ \citenamefont
  {Jelezko}}]{Unden2016}%
  \BibitemOpen
  \bibfield  {author} {\bibinfo {author} {\bibfnamefont {T.}~\bibnamefont
  {Unden}}, \bibinfo {author} {\bibfnamefont {P.}~\bibnamefont
  {Balasubramanian}}, \bibinfo {author} {\bibfnamefont {D.}~\bibnamefont
  {Louzon}}, \bibinfo {author} {\bibfnamefont {Y.}~\bibnamefont {Vinkler}},
  \bibinfo {author} {\bibfnamefont {M.~B.}\ \bibnamefont {Plenio}}, \bibinfo
  {author} {\bibfnamefont {M.}~\bibnamefont {Markham}}, \bibinfo {author}
  {\bibfnamefont {D.}~\bibnamefont {Twitchen}}, \bibinfo {author}
  {\bibfnamefont {A.}~\bibnamefont {Stacey}}, \bibinfo {author} {\bibfnamefont
  {I.}~\bibnamefont {Lovchinsky}}, \bibinfo {author} {\bibfnamefont {A.~O.}\
  \bibnamefont {Sushkov}}, \bibinfo {author} {\bibfnamefont {M.~D.}\
  \bibnamefont {Lukin}}, \bibinfo {author} {\bibfnamefont {A.}~\bibnamefont
  {Retzker}}, \bibinfo {author} {\bibfnamefont {B.}~\bibnamefont {Naydenov}},
  \bibinfo {author} {\bibfnamefont {L.~P.}\ \bibnamefont {McGuinness}}, \ and\
  \bibinfo {author} {\bibfnamefont {F.}~\bibnamefont {Jelezko}},\ }\href
  {\doibase 10.1103/PhysRevLett.116.230502} {\bibfield  {journal} {\bibinfo
  {journal} {Phys. Rev. Lett.}\ }\textbf {\bibinfo {volume} {116}},\ \bibinfo
  {pages} {230502} (\bibinfo {year} {2016})}\BibitemShut {NoStop}%
\bibitem [{\citenamefont {Reiter}\ \emph {et~al.}(2017)\citenamefont {Reiter},
  \citenamefont {S{\o}rensen}, \citenamefont {Zoller},\ and\ \citenamefont
  {Muschik}}]{Reiter2017}%
  \BibitemOpen
  \bibfield  {author} {\bibinfo {author} {\bibfnamefont {F.}~\bibnamefont
  {Reiter}}, \bibinfo {author} {\bibfnamefont {A.~S.}\ \bibnamefont
  {S{\o}rensen}}, \bibinfo {author} {\bibfnamefont {P.}~\bibnamefont {Zoller}},
  \ and\ \bibinfo {author} {\bibfnamefont {C.~A.}\ \bibnamefont {Muschik}},\
  }\href {\doibase 10.1038/s41467-017-01895-5} {\bibfield  {journal} {\bibinfo
  {journal} {Nat. Commun.}\ }\textbf {\bibinfo {volume} {8}},\ \bibinfo {pages}
  {1822} (\bibinfo {year} {2017})}\BibitemShut {NoStop}%
\bibitem [{\citenamefont {Sekatski}\ \emph {et~al.}(2017)\citenamefont
  {Sekatski}, \citenamefont {Skotiniotis}, \citenamefont
  {Ko{\l}ody{\'{n}}ski},\ and\ \citenamefont {D{\"{u}}r}}]{Sekatski2017a}%
  \BibitemOpen
  \bibfield  {author} {\bibinfo {author} {\bibfnamefont {P.}~\bibnamefont
  {Sekatski}}, \bibinfo {author} {\bibfnamefont {M.}~\bibnamefont
  {Skotiniotis}}, \bibinfo {author} {\bibfnamefont {J.}~\bibnamefont
  {Ko{\l}ody{\'{n}}ski}}, \ and\ \bibinfo {author} {\bibfnamefont
  {W.}~\bibnamefont {D{\"{u}}r}},\ }\href {\doibase 10.22331/q-2017-09-06-27}
  {\bibfield  {journal} {\bibinfo  {journal} {Quantum}\ }\textbf {\bibinfo
  {volume} {1}},\ \bibinfo {pages} {27} (\bibinfo {year} {2017})}\BibitemShut
  {NoStop}%
\bibitem [{\citenamefont {Demkowicz-Dobrza{\'{n}}ski}\ \emph
  {et~al.}(2017)\citenamefont {Demkowicz-Dobrza{\'{n}}ski}, \citenamefont
  {Czajkowski},\ and\ \citenamefont {Sekatski}}]{Demkowicz-Dobrzanski2017}%
  \BibitemOpen
  \bibfield  {author} {\bibinfo {author} {\bibfnamefont {R.}~\bibnamefont
  {Demkowicz-Dobrza{\'{n}}ski}}, \bibinfo {author} {\bibfnamefont
  {J.}~\bibnamefont {Czajkowski}}, \ and\ \bibinfo {author} {\bibfnamefont
  {P.}~\bibnamefont {Sekatski}},\ }\href {\doibase 10.1103/PhysRevX.7.041009}
  {\bibfield  {journal} {\bibinfo  {journal} {Phys. Rev. X}\ }\textbf {\bibinfo
  {volume} {7}},\ \bibinfo {pages} {041009} (\bibinfo {year}
  {2017})}\BibitemShut {NoStop}%
\bibitem [{\citenamefont {Zhou}\ \emph {et~al.}(2018)\citenamefont {Zhou},
  \citenamefont {Zhang}, \citenamefont {Preskill},\ and\ \citenamefont
  {Jiang}}]{Zhou2018achieving}%
  \BibitemOpen
  \bibfield  {author} {\bibinfo {author} {\bibfnamefont {S.}~\bibnamefont
  {Zhou}}, \bibinfo {author} {\bibfnamefont {M.}~\bibnamefont {Zhang}},
  \bibinfo {author} {\bibfnamefont {J.}~\bibnamefont {Preskill}}, \ and\
  \bibinfo {author} {\bibfnamefont {L.}~\bibnamefont {Jiang}},\ }\href
  {\doibase 10.1038/s41467-017-02510-3} {\bibfield  {journal} {\bibinfo
  {journal} {Nat. Commun.}\ }\textbf {\bibinfo {volume} {9}},\ \bibinfo {pages}
  {78} (\bibinfo {year} {2018})}\BibitemShut {NoStop}%
\bibitem [{\citenamefont {Layden}\ and\ \citenamefont
  {Cappellaro}(2018)}]{Layden2018}%
  \BibitemOpen
  \bibfield  {author} {\bibinfo {author} {\bibfnamefont {D.}~\bibnamefont
  {Layden}}\ and\ \bibinfo {author} {\bibfnamefont {P.}~\bibnamefont
  {Cappellaro}},\ }\href {\doibase 10.1038/s41534-018-0082-2} {\bibfield
  {journal} {\bibinfo  {journal} {npj Quantum Inf.}\ }\textbf {\bibinfo
  {volume} {4}},\ \bibinfo {pages} {1} (\bibinfo {year} {2018})}\BibitemShut
  {NoStop}%
\bibitem [{\citenamefont {Layden}\ \emph {et~al.}(2019)\citenamefont {Layden},
  \citenamefont {Zhou}, \citenamefont {Cappellaro},\ and\ \citenamefont
  {Jiang}}]{Layden2019}%
  \BibitemOpen
  \bibfield  {author} {\bibinfo {author} {\bibfnamefont {D.}~\bibnamefont
  {Layden}}, \bibinfo {author} {\bibfnamefont {S.}~\bibnamefont {Zhou}},
  \bibinfo {author} {\bibfnamefont {P.}~\bibnamefont {Cappellaro}}, \ and\
  \bibinfo {author} {\bibfnamefont {L.}~\bibnamefont {Jiang}},\ }\href
  {\doibase 10.1103/PhysRevLett.122.040502} {\bibfield  {journal} {\bibinfo
  {journal} {Phys. Rev. Lett.}\ }\textbf {\bibinfo {volume} {122}},\ \bibinfo
  {pages} {040502} (\bibinfo {year} {2019})}\BibitemShut {NoStop}%
\bibitem [{\citenamefont {Cappellaro}\ \emph {et~al.}(2019)\citenamefont
  {Cappellaro}, \citenamefont {Layden}, \citenamefont {Jiang}, \citenamefont
  {Zhou}, \citenamefont {Zhang},\ and\ \citenamefont
  {Preskill}}]{Cappellaro2019}%
  \BibitemOpen
  \bibfield  {author} {\bibinfo {author} {\bibfnamefont {P.}~\bibnamefont
  {Cappellaro}}, \bibinfo {author} {\bibfnamefont {D.}~\bibnamefont {Layden}},
  \bibinfo {author} {\bibfnamefont {L.}~\bibnamefont {Jiang}}, \bibinfo
  {author} {\bibfnamefont {S.}~\bibnamefont {Zhou}}, \bibinfo {author}
  {\bibfnamefont {M.}~\bibnamefont {Zhang}}, \ and\ \bibinfo {author}
  {\bibfnamefont {J.}~\bibnamefont {Preskill}},\ }in\ \href {\doibase
  10.1117/12.2511587} {\emph {\bibinfo {booktitle} {Optical, Opto-Atomic, and
  Entanglement-Enhanced Precision Metrology}}},\ \bibinfo {series and number}
  {\bibinfo {number} {March}},\ \bibinfo {editor} {edited by\ \bibinfo {editor}
  {\bibfnamefont {S.~M.}\ \bibnamefont {Shahriar}}\ and\ \bibinfo {editor}
  {\bibfnamefont {J.}~\bibnamefont {Scheuer}}}\ (\bibinfo  {publisher} {SPIE},\
  \bibinfo {year} {2019})\ p.~\bibinfo {pages} {51}\BibitemShut {NoStop}%
\bibitem [{\citenamefont {Nielsen}\ and\ \citenamefont
  {Chuang}(2010)}]{nielsen2010quantum}%
  \BibitemOpen
  \bibfield  {author} {\bibinfo {author} {\bibfnamefont {M.~A.}\ \bibnamefont
  {Nielsen}}\ and\ \bibinfo {author} {\bibfnamefont {I.~L.}\ \bibnamefont
  {Chuang}},\ }\href@noop {} {\emph {\bibinfo {title} {{Quantum computation and
  quantum information}}}},\ \bibinfo {edition} {10th}\ ed.\ (\bibinfo
  {publisher} {Cambridge University Press},\ \bibinfo {year}
  {2010})\BibitemShut {NoStop}%
\bibitem [{\citenamefont {Matsuzaki}\ \emph {et~al.}(2018)\citenamefont
  {Matsuzaki}, \citenamefont {Benjamin}, \citenamefont {Nakayama},
  \citenamefont {Saito},\ and\ \citenamefont {Munro}}]{Matsuzaki2018}%
  \BibitemOpen
  \bibfield  {author} {\bibinfo {author} {\bibfnamefont {Y.}~\bibnamefont
  {Matsuzaki}}, \bibinfo {author} {\bibfnamefont {S.}~\bibnamefont {Benjamin}},
  \bibinfo {author} {\bibfnamefont {S.}~\bibnamefont {Nakayama}}, \bibinfo
  {author} {\bibfnamefont {S.}~\bibnamefont {Saito}}, \ and\ \bibinfo {author}
  {\bibfnamefont {W.~J.}\ \bibnamefont {Munro}},\ }\href {\doibase
  10.1103/PhysRevLett.120.140501} {\bibfield  {journal} {\bibinfo  {journal}
  {Phys. Rev. Lett.}\ }\textbf {\bibinfo {volume} {120}},\ \bibinfo {pages}
  {140501} (\bibinfo {year} {2018})}\BibitemShut {NoStop}%
\bibitem [{\citenamefont {Maze}\ \emph {et~al.}(2008)\citenamefont {Maze},
  \citenamefont {Stanwix}, \citenamefont {Hodges}, \citenamefont {Hong},
  \citenamefont {Taylor}, \citenamefont {Cappellaro}, \citenamefont {Jiang},
  \citenamefont {Dutt}, \citenamefont {Togan}, \citenamefont {Zibrov},
  \citenamefont {Yacoby}, \citenamefont {Walsworth},\ and\ \citenamefont
  {Lukin}}]{Maze2008}%
  \BibitemOpen
  \bibfield  {author} {\bibinfo {author} {\bibfnamefont {J.~R.}\ \bibnamefont
  {Maze}}, \bibinfo {author} {\bibfnamefont {P.~L.}\ \bibnamefont {Stanwix}},
  \bibinfo {author} {\bibfnamefont {J.~S.}\ \bibnamefont {Hodges}}, \bibinfo
  {author} {\bibfnamefont {S.}~\bibnamefont {Hong}}, \bibinfo {author}
  {\bibfnamefont {J.~M.}\ \bibnamefont {Taylor}}, \bibinfo {author}
  {\bibfnamefont {P.}~\bibnamefont {Cappellaro}}, \bibinfo {author}
  {\bibfnamefont {L.}~\bibnamefont {Jiang}}, \bibinfo {author} {\bibfnamefont
  {M.~V.~G.}\ \bibnamefont {Dutt}}, \bibinfo {author} {\bibfnamefont
  {E.}~\bibnamefont {Togan}}, \bibinfo {author} {\bibfnamefont {A.~S.}\
  \bibnamefont {Zibrov}}, \bibinfo {author} {\bibfnamefont {A.}~\bibnamefont
  {Yacoby}}, \bibinfo {author} {\bibfnamefont {R.~L.}\ \bibnamefont
  {Walsworth}}, \ and\ \bibinfo {author} {\bibfnamefont {M.~D.}\ \bibnamefont
  {Lukin}},\ }\href {\doibase 10.1038/nature07279} {\bibfield  {journal}
  {\bibinfo  {journal} {Nature}\ }\textbf {\bibinfo {volume} {455}},\ \bibinfo
  {pages} {644} (\bibinfo {year} {2008})}\BibitemShut {NoStop}%
\bibitem [{\citenamefont {Liu}\ \emph {et~al.}(2015)\citenamefont {Liu},
  \citenamefont {Zhang}, \citenamefont {Chang}, \citenamefont {Yue},
  \citenamefont {Fan},\ and\ \citenamefont {Pan}}]{Liu2015}%
  \BibitemOpen
  \bibfield  {author} {\bibinfo {author} {\bibfnamefont {G.~Q.}\ \bibnamefont
  {Liu}}, \bibinfo {author} {\bibfnamefont {Y.~R.}\ \bibnamefont {Zhang}},
  \bibinfo {author} {\bibfnamefont {Y.~C.}\ \bibnamefont {Chang}}, \bibinfo
  {author} {\bibfnamefont {J.~D.}\ \bibnamefont {Yue}}, \bibinfo {author}
  {\bibfnamefont {H.}~\bibnamefont {Fan}}, \ and\ \bibinfo {author}
  {\bibfnamefont {X.~Y.}\ \bibnamefont {Pan}},\ }\href {\doibase
  10.1038/ncomms7726} {\bibfield  {journal} {\bibinfo  {journal} {Nat.
  Commun.}\ }\textbf {\bibinfo {volume} {6}},\ \bibinfo {pages} {1} (\bibinfo
  {year} {2015})}\BibitemShut {NoStop}%
\bibitem [{\citenamefont {Naghiloo}\ \emph {et~al.}(2017)\citenamefont
  {Naghiloo}, \citenamefont {Jordan},\ and\ \citenamefont
  {Murch}}]{Naghiloo2017}%
  \BibitemOpen
  \bibfield  {author} {\bibinfo {author} {\bibfnamefont {M.}~\bibnamefont
  {Naghiloo}}, \bibinfo {author} {\bibfnamefont {A.~N.}\ \bibnamefont
  {Jordan}}, \ and\ \bibinfo {author} {\bibfnamefont {K.~W.}\ \bibnamefont
  {Murch}},\ }\href {\doibase 10.1103/PhysRevLett.119.180801} {\bibfield
  {journal} {\bibinfo  {journal} {Phys. Rev. Lett.}\ }\textbf {\bibinfo
  {volume} {119}},\ \bibinfo {pages} {1} (\bibinfo {year} {2017})}\BibitemShut
  {NoStop}%
\bibitem [{\citenamefont {Wilde}(2013)}]{wilde2013quantum}%
  \BibitemOpen
  \bibfield  {author} {\bibinfo {author} {\bibfnamefont {M.~M.}\ \bibnamefont
  {Wilde}},\ }\href@noop {} {\emph {\bibinfo {title} {Quantum Information
  Theory}}}\ (\bibinfo  {publisher} {Cambridge University Press},\ \bibinfo
  {year} {2013})\BibitemShut {NoStop}%
\bibitem [{\citenamefont {Ekert}\ and\ \citenamefont
  {Macchiavello}(1996)}]{Ekert1996}%
  \BibitemOpen
  \bibfield  {author} {\bibinfo {author} {\bibfnamefont {A.}~\bibnamefont
  {Ekert}}\ and\ \bibinfo {author} {\bibfnamefont {C.}~\bibnamefont
  {Macchiavello}},\ }\href {\doibase 10.1103/PhysRevLett.77.2585} {\bibfield
  {journal} {\bibinfo  {journal} {Phys. Rev. Lett.}\ }\textbf {\bibinfo
  {volume} {77}},\ \bibinfo {pages} {2585} (\bibinfo {year}
  {1996})}\BibitemShut {NoStop}%
\bibitem [{\citenamefont {Bennett}\ \emph {et~al.}(1996)\citenamefont
  {Bennett}, \citenamefont {DiVincenzo}, \citenamefont {Smolin},\ and\
  \citenamefont {Wootters}}]{bennett1996mixed-state}%
  \BibitemOpen
  \bibfield  {author} {\bibinfo {author} {\bibfnamefont {C.~H.}\ \bibnamefont
  {Bennett}}, \bibinfo {author} {\bibfnamefont {D.~P.}\ \bibnamefont
  {DiVincenzo}}, \bibinfo {author} {\bibfnamefont {J.~A.}\ \bibnamefont
  {Smolin}}, \ and\ \bibinfo {author} {\bibfnamefont {W.~K.}\ \bibnamefont
  {Wootters}},\ }\href {\doibase 10.1103/PhysRevA.54.3824} {\bibfield
  {journal} {\bibinfo  {journal} {Phys. Rev. A}\ }\textbf {\bibinfo {volume}
  {54}},\ \bibinfo {pages} {3824} (\bibinfo {year} {1996})}\BibitemShut
  {NoStop}%
\bibitem [{\citenamefont {Knill}\ and\ \citenamefont
  {Laflamme}(1997)}]{Knill1997}%
  \BibitemOpen
  \bibfield  {author} {\bibinfo {author} {\bibfnamefont {E.}~\bibnamefont
  {Knill}}\ and\ \bibinfo {author} {\bibfnamefont {R.}~\bibnamefont
  {Laflamme}},\ }\href {\doibase 10.1103/PhysRevA.55.900} {\bibfield  {journal}
  {\bibinfo  {journal} {Phys. Rev. A}\ }\textbf {\bibinfo {volume} {55}},\
  \bibinfo {pages} {900} (\bibinfo {year} {1997})}\BibitemShut {NoStop}%
\end{thebibliography}%
\end{document}